# Buy Now, Pay Later: Academic Insights and Open Policy Questions

**Benedict Guttman-Kenney, Assistant Professor of Finance at Rice University**

**Walter W. Zhang, Assistant Professor of Marketing at The Wharton School, University of Pennsylvania**

**May 2026**

## I. Introduction

Buy Now, Pay Later (BNPL) has moved from a novelty product to mainstream consumer finance in the space of a few years. Before 2020, BNPL was a niche offering at the checkouts of online fashion merchants. BNPL then experienced substantial growth (Consumer Financial Protection Bureau, 2022; Salem & Udis, 2025), coinciding with increased online shopping since the onset of the COVID-19 pandemic in 2020. BNPL is now a pervasive payment option. In 2026, you can use BNPL to delay and split up payments for anything ranging from a pizza to your rent, and many items in between.

As the market has grown, so has the academic literature. There is now sufficient research into BNPL to take stock and review what we have learned. In this article, we summarize insights from the economics, finance, and marketing academic literature. While we have learned a great deal about BNPL in the last few years, we also discuss many remaining open questions and current concerns impacting policy decisions.

We start by explaining what BNPL products are and the business models of BNPL lenders. Then, we summarize the economics and psychology behind why some consumers use these products. Next, we describe the customer base that uses BNPL. We show the effects of BNPL on consumers, an area of research with substantial policy interest. Another area of policy interest is whether BNPL loans should be reported in a consumer's credit report. We explain why BNPL is typically not on credit reports and its implications. Finally, we offer some concluding thoughts.

## II. What is BNPL?

BNPL is a short-term loan or installment credit product, typically offered at the point of sale and at online checkout. Popular BNPL lenders are Affirm, Afterpay, Klarna, PayPal, Sezzle, and Zip. The standard structure of a BNPL loan involves splitting a purchase into a small number of equal payments, often three or four, repaid over weeks or months. The loans usually have no interest and no fee if repayments are made on time. Late fees, where they apply, are generally capped and modest compared to the penalty charges or compounding interest common in credit cards or overdrafts.

The BNPL market encompasses several distinct product types, a distinction the Consumer Financial Protection Bureau (2022) drew in its investigation into this market. The most prominent product is the "pay-in-four" model, where repayments are split into four equal installments charged to a debit or credit card. Though some products, such as Klarna's "Pay in 30 days" product, offer consumers the ability to buy now and pay 30 days later without cost, somewhat conceptually similar to an interest-free grace period on a credit card. Longer-term BNPL products, sometimes carrying interest, occupy a different part of the market and naturally carry different risk profiles.

The literature does not always distinguish clearly between these variants, leaving open whether some BNPL loan structures are better for consumers, or more profitable to lenders, than others. Traditional lenders also increasingly offer BNPL-type products, enabling consumers to break up transactions on their credit card or their debit card, seemingly as a competitive response to BNPL's growth (Alcazar & Bradford, 2021). However, we do not know much about these beyond the research of Donnelly et al. (2024).

BNPL is not an isolated development; it is one of several FinTech lending innovations to emerge, using technological advances to offer consumers increased convenience and speed (Berg, Fuster, & Puri, 2022). Bian, Cong, and Ji (2024) and Wang (2025) place BNPL in a broader landscape of payment innovations, situating it alongside the recent growth in super-app e-wallets and digital payments driven by FinTech innovations. The emergence of BNPL as both a form of lending and a means of payment makes it, and also other FinTech lending and payment innovations, challenging to regulate.

## III. BNPL Business Model

To understand BNPL's incentives and risks, it helps to understand how BNPL providers make money. The BNPL model is structurally different from traditional consumer credit, and that difference has significant implications for the alignment of provider and consumer interests.

BNPL providers typically do not charge consumers interest, at least on the most used BNPL products. Some BNPL providers also do not charge any fees to consumers, in contrast to most other credit products. Instead, revenue derives primarily from fees charged to merchants: BNPL providers take a percentage of each transaction, often notably higher than standard debit or credit card interchange fees. The logic of this BNPL model is that offering BNPL to consumers increases conversion rates at checkout and average order values, delivering enough additional revenue to justify merchants paying the higher fees.

Berg et al. (2025) examine the economics of BNPL from the merchant's perspective in detail. They find that a merchant's willingness to pay BNPL fees depends on the lift in sales the product generates. While the lift is real, its magnitude is likely to be context dependent. BNPL enables merchants to price discriminate between consumers with different willingness-to-pay for a product. Merchants bundle a product with an interest-free BNPL loan, enabling consumers with a lower-willingness-to-pay for the underlying product, who are also more likely to be less creditworthy, to purchase the product.

Alok, Ghosh, Kalda, and Mitra (2026) examine merchant heterogeneity in the BNPL market, documenting that the competitive dynamics and benefits of BNPL differ substantially across types of merchants, with implications for market concentration and the distribution of gains between providers and merchants.

Desai and Jindal (2024) show that BNPL is not just a payment option but a competitive retail strategy: it can expand demand, raise prices, and change how firms compete. While BNPL benefits a monopolist, in competitive markets either one or more merchants may offer it depending on spillovers and pricing incentives. Consumers may underestimate BNPL costs, which may lead to higher prices, overbuying, and costly upgrades. Further research can help us to understand how different levels of retail competition interact with BNPL to affect consumer behavior.

## IV. The Economics & Psychology of Why Consumers Use BNPL

The surface-level appeal of BNPL is obvious: interest-free credit seamlessly integrated at the point of sale is an attractive proposition to consumers. Economic theory suggests that consumers should value the opportunity to smooth out purchases interest-free, as a dollar today is worth more than a dollar tomorrow. By using BNPL, consumers are discounting the economic cost of purchasing a good. Separately, some consumers may simply prefer BNPL as a convenient method of payment over other alternatives.

The economic benefits of using BNPL can be larger if consumers have binding liquidity constraints. This means some consumers can afford the product, but do not have enough cash or credit available right now to pay for it. Consumers may lack cash for many reasons. For example, their paycheck may not have arrived yet, they may have had an unexpected bill, or they experienced a temporary drop in income. Without BNPL (or some other credit product), a liquidity-constrained consumer may not be able to purchase a product, whereas they can potentially do so with BNPL.

The economic value of BNPL may be especially important in the context of ecommerce, where BNPL demand is especially strong. For clothing stores, a consumer may be unsure about which product will fit them or look good on them. A consumer may want to order a bundle of clothes in different sizes and styles to try on in their own home, with the intention to return the ones they do not want. A liquidity-constrained consumer may not be able to make such purchasing decisions, as they cannot commit their precious cash to this purchase for weeks, which may be the time to receive a refund from returns. BNPL or other credit products (such as the grace period on a credit card) enable consumers with limited cash to access a broader set of products.

However, there is reason to think that some of the consumer demand for BNPL is not only driven by traditional economic factors, but also by psychological factors. BNPL decouples the pleasure of purchasing a good today from the pain of paying for it, deferring that painful cost until later. Part of the reason why consumers may use BNPL is that they cannot resist doing so. If so, offering BNPL may help to facilitate impulsive spending activity by consumers. Unfortunately, BNPL may be especially attractive to consumers who lack self-control or financial sophistication, or those who are overconfident in their ability to repay such debt. If BNPL is driving increased spending, that may drive economic growth, however, such growth may not necessarily be sustainable, potentially later leading to problems for consumers, lenders, or merchants that rely on BNPL.

## V. Who Uses BNPL?

Understanding who uses BNPL matters for assessing its effects. A product used mainly by financially stable consumers as a convenience tool has different welfare implications than one used heavily by consumers who are already liquidity constrained.

The evidence consistently suggests that BNPL use is not evenly distributed across the population. Using survey data linked to credit bureau records, Shupe, Fulford, and Li (2023) find that BNPL borrowers are more likely than non-users to show markers of financial stress, including higher indebtedness, credit card statement balances, delinquencies, and lower credit scores. Similarly, Shupe and DeLuca (2025) show that BNPL borrowers often also carry other unsecured debt. Akana and Zeballos Doubinko (2023) show that credit bureau data does not predict BNPL usage. Stavins (2024) finds that BNPL use is higher among financially vulnerable consumers, broadly consistent with the findings of research by Di Maggio, Williams, and Katz (2022) and deHaan et al. (2024).

Demand for BNPL points in the same direction. Aidala et al. (2026) find that demand for standard BNPL is higher among younger, lower-income, and less creditworthy consumers, consistent with evidence from Ireland (Jose and Kelly, 2025). Financial vulnerability is only part of the story. Larrimore, Lloro, Merchant, and Tranfaglia (2025) show that BNPL serves different functions for different households: many users describe BNPL as a convenient way to spread payments, while many households also report using BNPL because it was the only way they could afford the purchase. Wheat et al. (2026) make a similar distinction in the homeowner context, framing BNPL as either a convenience tool or a liquidity valve when household balance sheets are under strain. A survey by the Board of Governors of the Federal Reserve System (2025) reinforces the point that BNPL is now part of ordinary household cash-flow management, while also documenting late payment among a growing share of users over time.

This heterogeneity in how consumers use BNPL creates a challenge for policymakers. For some consumers, BNPL may be a low-cost way to smooth spending without using more expensive credit and alleviates financial distress. For others, especially those already juggling debt or with strained finances, BNPL simply adds another layer of debt, further increasing their financial distress.

Payment design also matters. Liu, Zhang, and Zou (2026) study the mirror-image "Pay Now, Buy Later" model and show how consumer-facing payment products can be marketed as win-win arrangements while generating profits from inattention, unused balances, and misaligned firm incentives. BNPL users are heterogeneous: many use it for convenience, while many others use it because they are financially constrained. That makes it hard to distinguish helpful cash-flow smoothing from risky debt layering, especially when BNPL obligations are not consistently visible to other lenders.

## VI. Effects of BNPL on Consumers

The central welfare question is not simply whether BNPL expands access to credit. It is whether that access improves consumers' ability to manage temporary cash-flow constraints, or instead encourages additional borrowing by consumers who are already financially stretched. The emerging evidence points in both directions.

Guttman-Kenney, Firth, and Gathergood (2023) provide one of the first academic studies of BNPL, using UK credit card transaction data to track consumer use of BNPL. They find that some consumers pay BNPL obligations with credit cards, turning what begins as a short-term, interest-free installment loan into ordinary credit card debt that may carry high interest rates.

deHaan et al. (2024) find that new BNPL users experience increases in overdraft charges and credit card interest and fees relative to non-users. In contrast, Chan et al. (2025) study UK regulatory data on BNPL use linked to credit files and find no strong evidence that BNPL increases broader consumer indebtedness or missed payments.

Other evidence is more mixed: Papich (2022) finds that access to BNPL increases borrowing, but also reports improvements in some repayment outcomes, including lower past-due balances and fewer current delinquencies. Shupe and Palloni (2025) similarly examine whether BNPL affects consumers' ability to repay non-BNPL debts, finding little evidence that BNPL changes repayment behaviors on those obligations. The effects depend heavily on the borrower's financial condition and interaction with their other debts. For Australia, Gerrans, Baur, and Lavagna-Slater (2022) frame this as a responsibility problem: Although BNPL can be cheaper than credit cards when used carefully, its product design and lighter regulatory treatment raise concerns about whether providers have sufficient incentives to prevent consumer harm.

The spending effects are clearer. Di Maggio, Williams, and Katz (2022) find that BNPL access increases total spending, including among liquidity-constrained consumers. Their effect sizes are too large to be explained by a traditional economic explanation based around liquidity constraints. Instead, their results reflect a behavioral story in which BNPL has a psychological impact on consumers, shifting their expenditures toward particular goods.

Product design and marketing may amplify the risks to consumers. Burg and Keil (2026) find that BNPL availability can increase impulse purchases, while Kumar, Salo, and Bezawada (2024) and Maesen and Ang (2025) show that BNPL alters online purchase behavior, including purchase activity and purchase amounts. Aggarwal, Kaye, and Odinet (2022) add that social media channels, such as #FinTok on TikTok, have become important spaces for consumer-finance promotion and discussion, making BNPL harder to monitor via advertising rules designed for traditional channels.

All in all, BNPL is neither clearly good nor clearly bad for consumers. The effects of BNPL vary across consumers and appear domain-specific. BNPL can help some consumers with short-term cash-flow problems, but it can also encourage extra spending and increase financial distress for some already-stretched consumers. This concern becomes sharper when BNPL debts are not visible in credit reports.

## VII. BNPL Credit Reporting

Arguably the largest open issue surrounding BNPL concerns credit reporting. BNPL transactions are not consistently reported to credit bureaus in the United States, so using BNPL does not build your credit history or improve your credit score. Most BNPL lenders do not report BNPL loans taken out, although loans that consumers miss payments on may be reported. This is a problem for both BNPL and non-BNPL lenders as it means that a lender does not know how much debt a consumer has at the time of applying for a loan. One concern is that this enables "loan stacking" where one consumer can take out BNPL loans from multiple BNPL lenders at one time, a behavior that may be risky for both the borrower and their lenders.

While this piece focuses on BNPL lenders, not reporting loans to credit bureaus is not a BNPL-specific issue. Some non-BNPL lenders (e.g., payday lenders and some subprime auto lenders) also do not report their data to credit bureaus. Even lenders that do report to credit bureaus do so selectively. For example, Guttman-Kenney & Shahidinejad (2025) show that none of the largest credit card lenders report to the credit bureaus how much their cardholders pay against their credit card bill.

It may appear surprising that not all debts held by a consumer are on their credit report, but that is the system in the United States, where no rule requires any lender to report their information to a credit bureau (Gibbs et al., 2025). Instead, each lender decides whether to voluntarily share information with one or more credit bureaus (Equifax, Experian, and TransUnion), after weighing whether it is privately optimal for them to do so, given the costs and benefits. This creates a potential market failure as positive externalities of more information that improves other lenders' decisions are not taken into account. There is also non-reciprocity in this system: if one lender shares information with a credit bureau, another lender can observe such information even if they do not share information on their own customers (Guttman-Kenney & Shahidinejad, 2025). This system raises the competitive risk to one lender unilaterally sharing information.

Broadly there are two reasons why BNPL lenders do not share information with credit bureaus. The first reason is that BNPL lenders are concerned that reporting information would adversely affect their customers' credit scores, which would reduce customers' demand for BNPL.

Why are they worried about this? Historically, the main credit score that lenders rely on is a FICO score, though VantageScore, jointly owned by the credit bureaus, is a growing competitor. In these scores, opening or closing several accounts in quick succession reduces your credit score, because historically such behaviors are correlated with higher defaults. These scoring models do not distinguish between a BNPL loan and another form of unsecured personal loan, so reporting BNPL loans would be expected to often reduce consumers' FICO credit scores, as these loans are designed to be closed soon after opening. A simple adjustment of credit scores to account for BNPL loans being different from other forms of credit may also not help to improve credit scores of BNPL users if, on average, taking out a BNPL loan is correlated with a consumer being at higher risk of not repaying their loans.

Given this context, it is important to understand the informativeness of BNPL use as a credit risk signal. The only academic study of this is Laudenbach et al. (2025), who find that BNPL use is a positive signal of creditworthiness in their Norwegian data. We do not know if this finding translates to the United States.

In the United States, Affirm and FICO have done some work to try to correct this technical issue and have developed a version of FICO that accounts for BNPL, however, most lenders rely on older versions of FICO scores that would not do so. Crucially, other BNPL lenders have not shared data, so the Affirm-FICO solution, so far, has not satisfied other BNPL lenders, who have different customer bases and products, and so their customers may be differentially impacted.

The second, more fundamental, reason why BNPL lenders do not share information with credit bureaus is due to competition concerns. In the United States, credit reporting data is not only used for assessing a credit applicant's creditworthiness. Lenders also use credit reporting data to target their marketing to acquire new customers. BNPL lenders are concerned that reporting information would enable competing BNPL lenders and non-BNPL lenders to poach a BNPL lender's profitable customers, after the BNPL lender's technology had done the initial hard work of assessing a customer's risk. Given this fundamental issue, it appears unlikely that BNPL lenders will start voluntarily reporting information, unless there is regulatory pressure to do so. One potential insight into what might happen if firms are forced to share data is the study of Guttman-Kenney & Shahidinejad (2025), which shows that when the Federal Trade Commission required credit card lenders to report information on their credit card's credit limits, something that was not privately optimal for them to do, it increased switching, consistent with increased competition.

Understanding the domains around the world where BNPL lenders are willing to share information can also help to understand why they do not do so in the United States. For example, BNPL lenders share information with credit bureaus in the United Kingdom. Why are they willing to do so there? Partially pressure from regulators but also access to credit reporting data is only on a reciprocal basis and is not used for marketing. Therefore, there is limited competitive threat of a BNPL lender's competitors targeting their profitable customers if a BNPL lender shares information in the United Kingdom, in sharp contrast to the United States.

The academic research does not tell us the effects of mandating that BNPL lenders share information. From a competition perspective, we might be worried about requiring BNPL lenders to report, without requiring non-BNPL lenders to also do so. Studying the effects of BNPL being reported in other parts of the world can be informative. Doing so can enable learning about which consumers would benefit and lose out from information being shared, and whether sharing information is net beneficial. Important open questions include: How does reporting BNPL debt change consumer demand for BNPL? How does reporting change the decisions of non-BNPL lenders? How does reporting affect competition between BNPL and non-BNPL lenders?

There is limited discussion on whether the current structure of credit reporting can be improved upon. For example, if consumers increasingly use short-term credit products, such as BNPL, a more real-time credit reporting system may appear more suitable to prevent loan stacking than the current historic approach that is largely based on a monthly reporting cycle.

More broadly, the credit system in the United States traditionally requires consumers to take out credit to build credit. Credit cards are a key part of that, often being the first product a consumer takes out. Yet, there is substantial research showing many consumers get into financial difficulties with credit cards, with lenders being highly sophisticated in the design of their product features and targeted marketing of such products to lead consumers to make costly mistakes. In theory, BNPL could potentially offer a lower-risk way for consumers to demonstrate their creditworthiness and build their credit scores.

## VIII. Concluding Thoughts

The BNPL literature has grown rapidly, and regulatory frameworks have struggled to keep pace. This is not least because BNPL's hybrid nature sits awkwardly within categories designed for either payments or credit. An interpretive rule issued by the Consumer Financial Protection Bureau in 2024 to regulate BNPL similar to credit cards was reversed by the Trump administration a year later (see Congressional Research Service, 2026, and Wang, 2025, for more discussion of such policy issues).

There remains an open question on how to design disclosures for the digital and AI world. The existing disclosure regimes did not have products such as BNPL or digital interfaces in mind. More broadly, the role of celebrity influencers and non-traditional forms of advertising driving consumer spending and use of financial products is increasingly important to assess and challenging to regulate.

There remains substantial scope to better understand the nature of competition between BNPL lenders, and also between BNPL and traditional credit products. When merchants offer multiple payment options, which BNPL option do consumers choose and why? What determines which payment method consumers use, and whether BNPL will replace credit cards remains under-studied. While focus has been on BNPL, other FinTech lending products have emerged, with much less academic scrutiny. For example, there is limited research (Alcazar & Bradford, 2020; Murillo, Vallée, & Yu, 2025) on Earned Wage Access (EWA) credit products that appear to be increasingly offered directly to consumers or via partnerships with employers. The academic research has entirely focused on the business-to-consumer BNPL, however, business-to-business BNPL can potentially be just as economically important and disrupt traditional slow, manual trade credit.

There is an established demand for BNPL products from consumers. However, such demand has not been reflected in the stock market performance of BNPL lenders. BNPL lenders' prospects appear less rosy recently. An important open question is whether the BNPL business model is sustainable. Will merchants still be willing to pay for BNPL if regulation or innovation reduces the cost of other payment methods? Can BNPL only be viable as one part of a broader portfolio of product offerings to consumers and businesses? Ultimately, whether the BNPL business model is robust to a severe recession that sharply reduces spending and increases defaults has yet to be tested.